\documentclass[namedreferences]{solarphysics}
\usepackage[optionalrh]{spr-sola-addons} % For Solar Physics 
\usepackage{graphicx}        % For eps figures, newer & more powerfull
\usepackage{amsbsy}        % useful mathematical symbols
\usepackage{color}           % For color text: \color command
\usepackage{url}             % For breaking URLs easily trough lines
\usepackage{rotating}             % For the sideways figure
\usepackage[breaklinks,colorlinks,urlcolor=blue,citecolor=blue,linkcolor=blue]{hyperref}

\renewcommand{\vec}[1]{ {\mathbf #1} }

\newcommand{\Alfven}{{Alfv\'{e}n} }

\begin{document}

\begin{article}

\begin{opening}

\title{Interchange Reconnection \Alfven Wave Generation}

\author{B.J.~\surname{Lynch}$^{1}$\sep
        J.K.~\surname{Edmondson}$^{2}$\sep
        Y.~\surname{Li}$^{1}$
       }
\runningauthor{B. J. Lynch et al.}
\runningtitle{Interchange Reconnection Wave Generation}

   \institute{$^{1}$ Space Sciences Laboratory, University of California, Berkeley, CA 94720, USA 
                     email: \url{blynch@ssl.berkeley.edu}; \url{yanli@ssl.berkeley.edu} \\
	      $^{2}$ Atmospheric, Oceanic and Space Sciences Department, University of Michigan, Ann Arbor, MI
	             48109, USA email: \url{jkedmond@umich.edu} \\
             }

\begin{abstract}

Given recent observational results of interchange reconnection
processes in the solar corona and the theoretical development
of the S-Web model for the slow solar wind, we extend the analysis of the 3D MHD simulation of interchange reconnection by
\citeauthor{Edmondson2009} (\textit{Astrophys. J.} \textbf{707}, 1427,
\citeyear{Edmondson2009}). Specifically, we analyze the consequences
of the dynamic streamer-belt jump that corresponds to flux opening
by interchange reconnection.
Information about the magnetic field restructuring by interchange
reconnection is carried throughout the system by \Alfven waves
propagating away from the reconnection region, distributing the shear
and twist imparted by the driving flows, including shedding the injected
stress-energy and accumulated magnetic helicity along newly open fieldlines.
We quantify the properties of the reconnection-generated wave activity in
the simulation. There is a localized high-frequency component associated
with the current sheet/reconnection site and an extended low-frequency
component associated with the large-scale torsional \Alfven wave
generated from the interchange reconnection field restructuring. The
characteristic wavelengths of the torsional \Alfven wave reflect the
spatial size of the energized bipolar flux region. Lastly, we discuss avenues of future research
by modeling these interchange reconnection-driven waves and investigating
their observational signatures.

\end{abstract}


\end{opening}
%-------------------------------------------------

% ===================================================================
\section{Introduction}
% ===================================================================

The interchange reconnection (IR) model for the solar
wind was introduced by Fisk and co-authors to explain
in-situ observations of the interplanetary magnetic field
\cite{Fisk1999,Fisk2001,Fisk2005,Fisk2006} and various properties
of the in-situ slow solar wind, including its high variability, lateral
extent, elemental and ionic composition (e.g., \opencite{Geiss1995};
\opencite{Gosling1997}; \opencite{Zurbuchen2007}). 
\inlinecite{Antiochos2011} and colleagues developed
the \textsl{Separatrix Web} (S-Web) model to bridge the gap between
the quasi-steady and the IR solar wind models, incorporating relevant
aspects of dynamic IR behavior via self-consistent MHD modeling and
rigorous topological analysis of the solar corona's temporal evolution
(see \opencite{Linker2011}; \opencite{Titov2011}).

There has also been a long history of the development and refinement of wave-heating and turbulent dissipation models for the generation of the solar wind (e.g. \opencite{Hollweg1986}; \opencite{Matthaeus1999}; \opencite{Cranmer2007}; \opencite{Cranmer2010}; \opencite{Ofman2004}, \citeyear{Ofman2010}; \opencite{Verdini2010}). These tend to work pretty well for the fast solar wind emanating from coronal holes, and appear to be making progress towards reproducing some of the elemental and ionic composition properties of the slow wind (\opencite{Cranmer2007}; \opencite{Laming2004}, \citeyear{Laming2009}; \opencite{Bryans2009}). However, for the most part, the wave-heating models start with an initial wave power spectrum and concentrate on the transformation and evolution of this spectra, calculating the turbulent cascade to higher frequencies and ultimately the deposition of this energy into plasma heating and acceleration, typically through various wave-particle interactions or resonances with particular ions. 

%=====================

The purpose of this paper is to extend the analysis of the IR simulation presented
by \inlinecite{Edmondson2009} to characterize the wave activity that occurs as a consequence of the IR opening of
large, closed field-lines. We examining the material and energy fluxes at the open--closed field boundary and show that IR transports the twist component of the previously closed
flux into the open field in the form of a large-scale torsional \Alfven
wave.
While our simulation is highly idealized and the footpoint shearing
motions used to energize the configuration are not meant to model
observed photospheric flow patterns, we feel the large-scale vortical
boundary flow that imparts twist to the entire flux system is a convenient
representation of an aspect of the \inlinecite{Antiochos2013} helicity
condensation theory.

The \inlinecite{Antiochos2013} helicity condensation model explains the accumulation of shear and/or
twist at polarity inversion lines. They describe an inverse cascade
transportation process where small-scale twist emerges and reconnects
to larger and larger scales. This process is naturally limited by the
topological boundaries of the flux system; the large scale, accumulated
shear/twist condenses at both the polarity inversion line and the
separatrix boundary between open and closed field. While the low-lying,
highly sheared fields of filament channels are a common occurrence, we
do not observe highly sheared fields at the boundaries between open
and closed flux. \inlinecite{Antiochos2013} argued that the helicity
that tries to accumulate at the coronal hole boundaries is, in fact,
transferred onto open field-lines through IR processes occurring at the
open--closed interface, and thus regularly escapes the corona.
This is a reasonable conjecture since the interplanetary magnetic field
and its fluctuations are known to transport magnetic helicity (e.g.,
\citeauthor{Smith1999}, \citeyear{Smith1999} and references therein).

The structure of the article is as follows. In Section~\ref{S:Overview},
we present a brief overview of the MHD simulation and describe the
topological evolution of interchange reconnection, illustrate the
nonlinear, large-scale torsional \Alfven wave generation and quantify the
opening of previously closed flux. 
In Section~\ref{S:Outflow-Flux} we quantify the associated material
and energy fluxes. In Section~\ref{S:Waves} we examine the spatial extent and
radial propagation of the IR-generated torsional \Alfven wave fluctuations
through our computational domain and examine the spatial structure and
properties of the wave power spectra. In Section~\ref{S:Discussion}, we make suggestions for future observational
tests, and discuss improvements for future numerical simulation work in
this area.

% ===================================================================
\section{Overview of the MHD Simulation}
\label{S:Overview}
% ===================================================================

\subsection{Numerical Methods and Initial Conditions}
\label{S:Overview-MHD}

The \inlinecite{Edmondson2009} simulation was run with the Adaptively
Refined MHD Solver (ARMS) code, developed by C.~Richard~DeVore
and collaborators. 
ARMS calculates solutions to the 3D nonlinear, time-dependent MHD
equations that describe the evolution and transport of density,
momentum, and energy throughout the system, and the evolution of the
magnetic field and electric currents (see e.g., \opencite{DeVore2008}). 
The numerical scheme used is a finite-volume, multidimensional flux-corrected transport (FCT)
algorithm \cite{DeVore1991}. 
The ARMS code is fully integrated with the adaptive-mesh toolkit PARAMESH
\cite{MacNeice2000}, to handle dynamic, solution-adaptive grid refinement
and support an efficient multiprocessor parallelization.

\begin{figure}    %%%%%%%%%%%%%%%%%% FIGURE 1 
   \centerline{\includegraphics[width=1.0\textwidth,clip=]{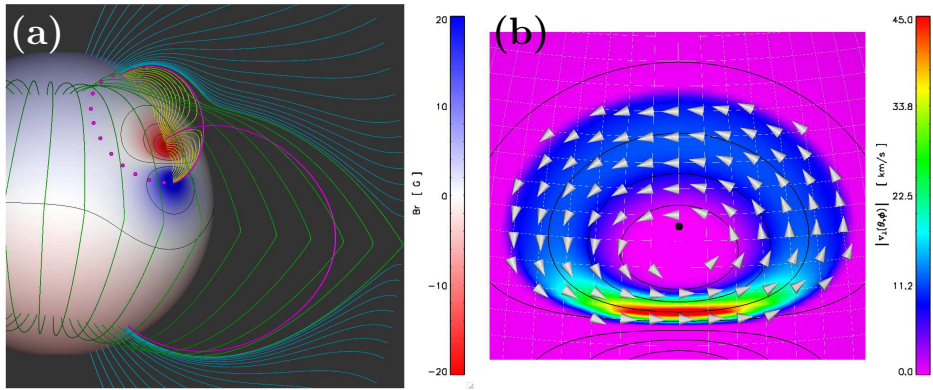}}
   \vspace{-0.41\textwidth}   % Shift close to the panel top 
   \centerline{\Large \bf     % Includes the labels (here needs the color 
                                %   package, see beginning of this file)
   \hspace{-0.010\textwidth}  \color{white}{(a)}
   \hspace{0.430\textwidth}  \color{black}{(b)} \hfill}
   \vspace{0.36\textwidth}    % Shift back to the panel bottom 

   \caption{Panel (a) initial potential field showing the active-region
   flux (yellow field lines), the closed field streamer-belt boundary
   (green field lines) and the open field (blue field lines). The AR
   separatrix boundary on the solar surface is indicated with magenta
   dots, and the magenta fieldlines show the separatrix dome and inner
   and outer spine field lines. Panel (b) velocity profile shearing
   pattern applied to the red (negative) polarity AR flux. Adapted from
   Edmondson \textit{et al.} (2009).
   }
   \label{F1}
\end{figure}

Figure~\ref{F1}(a) shows the initial magnetic-field configuration of
our system. This is the well-known 3D embedded bipole / null-point
potential field in spherical coordinates, the simplest nontrivial 3D
structure that facilitates magnetic reconnection in response to system
energization. The solar-surface contour map plots the radial component
of the magnetic field. There are two flux systems associated with the two
polarity-inversion lines separated by a hemispherical separatrix dome
where the continuous intersection with the solar surface is indicated
as the ring of magenta dots. The yellow field lines show the extent of
the AR flux system while the green field lines trace the boundary of
the streamer-belt flux system. Representative field lines of the open
field regions are drawn in blue. The inner and outer spine field lines
are shown in magenta as well as the boundary of the separatrix dome in
the $\phi=0$ plane. There is a 3D null point at the intersection of the
spine field lines with the separatrix done. Figure~\ref{F1}(b) plots the
maximum magnitude of the applied shearing flows in the negative polarity
spot of the AR (colored red in panel a). The flow field is constructed
to follow the contours of $B_r$ on the $r = R_\odot$ boundary so the
normal flux distribution remains constant throughout the simulation. The
shearing flow magnitude is roughly $6\%$ of the global average \Alfven
speed and is smoothly ramped up from $0 \le t \le$ 2~000~s, uniform from
2~000 $\le t \le$ 20~000~s, and ramped back down from 20~000 $\le t \le$
22~000~s. We refer to \inlinecite{Edmondson2009} for details
of the initial magnetic-field model, properties of the initial solar
atmosphere, the functional form of the shearing profile, and details of
the simulation boundary conditions.

%
%=====================
%

% ===================================================================
\subsection{Topological Evolution of Interchange Reconnection}
\label{S:Overview-Top}
% ===================================================================

The magnetic topology and evolution of the \inlinecite{Edmondson2009}
simulation is essentially a large, spherical version of the
twist-jet model (e.g., \opencite{Shibata1986}; \opencite{Pariat2009},
\citeyear{Pariat2010}) and is qualitatively similar to the flux emergence
scenario of \inlinecite{Torok2009}.
These authors have identified the outwardly propagating torsional
\Alfven waves resulting from the reconnection dynamics and have suggested their potential contribution to
coronal heating and the acceleration of the solar wind.
The placement of the parasitic polarity spot near a well-defined
open--closed field boundary (i.e., the edge of the helmet streamer belt)
allowed \inlinecite{Edmondson2009} to track the evolution of the open
flux and the open--closed boundary during the interchange reconnection
process and allows us to perform a quantitative analysis of the physical
properties of the simulation outflow originating at the edge of the streamer belt.

\begin{figure}   %%%%%%%%%%%%%%%%%% FIGURE 2 
   \centerline{\includegraphics[width=1.0\textwidth,clip=]{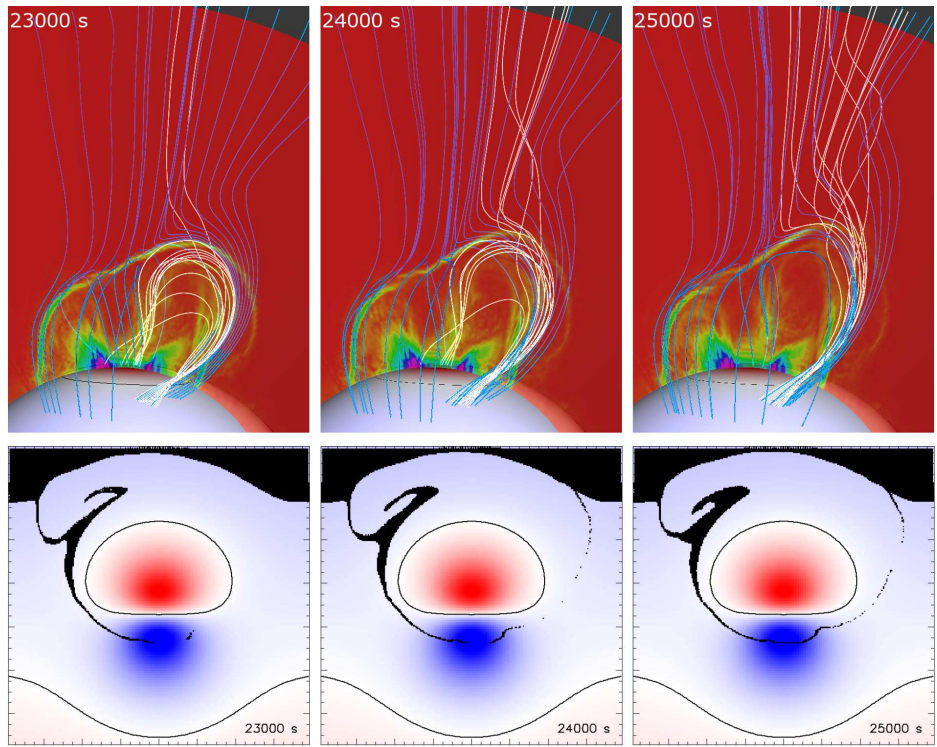}}
   \caption{Top row, \textit{Heliospace} renderings of the simulation data. The white
   field lines traced from $r=R_\odot$ footpoints are the same field lines in
   each panel against backdrop of current density magnitude. Bottom
   row, frames of $B_r(R_\odot,\theta,\phi)$ with the open flux area 
   area overplotted as the black pixel map showing the evolution of the
   narrow open field corridor that traces the separatrix surface
   of the AR flux system.}
   \label{FBipoleReconnection}
\end{figure}

The top row of Figure~\ref{FBipoleReconnection} illustrates both the
interchange reconnection process occurring at the stressed null point
and the propagation of the kinks and twist in the newly reconnected open
field lines (the white lines). The planar surface shows the magnitude
of current density, strongest in the core of the sheared active-region
flux, but also outlining the current sheet that has formed along the
separatrix boundary of the AR flux system. In each panel the white field
lines are plotted from the same (stationary) footpoints on the solar
surface so their evolution depicts the flux transfer of originally closed,
stressed AR field through the current sheet and into the open field region
where the stress and helicity (twist component) propagate towards the
outer-radial boundary. The bottom row of Figure~\ref{FBipoleReconnection}
shows the spatial distribution of the open flux footpoints on the
lower $r = R_\odot$ boundary at each corresponding time. The narrow,
open field corridor is in the process of opening up along the entire
separatrix boundary as the outer spine line becomes open (as discussed
in \opencite{Edmondson2009}).

\begin{figure}[t]    %%%%%%%%%%%%%%%%%% FIGURE 3 
   {\includegraphics[width=1.0\textwidth, clip]{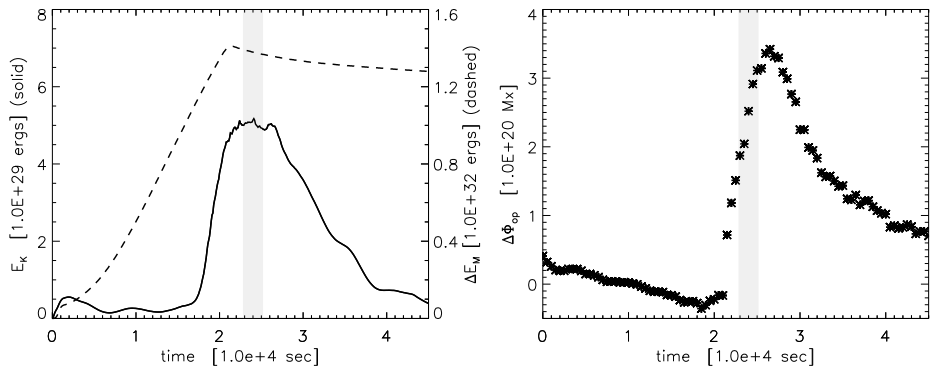}}
   \caption{Left panel: kinetic energy $E_{\rm K}$ (solid)
   and the free magnetic energy $\Delta E_{\rm M}$ (dashed) evolution
   during the simulation (adapted from Edmondson \textit{et al.},
   2009). Right panel: change in open flux $\Delta \Phi_{\rm
   op} = \Phi_{\rm op} - \langle \Phi_0 \rangle$ in a spherical
   subdomain.
   The light-gray bar indicates the time period of interchange reconnection that opens previously closed flux shown in
   Figure~\ref{FBipoleReconnection}.}
   \label{Feng}
\end{figure}

The global evolution of the magnetic and kinetic energies during the
simulation are shown in Figure~\ref{Feng}. The left panel plots the total
kinetic energy ($E_{\rm K}$, solid line) and the free magnetic energy
(dashed line) within the entire computational domain.  The light-gray bar
indicates the time interval shown in Figure~\ref{FBipoleReconnection}. We
define the change in magnetic energy from the initial, potential state
as $\Delta E_{\rm M} = E_{\rm M}(t) - E_{\rm M}(0)$ with $E_{\rm M}(0)=
2.597 \times 10^{33}$~erg. Since the normal magnetic-flux distribution
is fixed throughout the simulation, $\Delta E_{\rm M}$ represents the
free magnetic energy of the system above the initial potential state.
The amount of free magnetic energy released via the IR process ($\sim$10$^{31}$~erg) is only
$\sim$10\% of the maximum stored value. In terms
of the total free energy in the system, a relatively small amount of magnetic
energy release can result in significant topological restructuring and the 
wave generation we will examine herein.

The right panel of Figure~\ref{Feng} plots the evolution of
the open flux. We calculated the open flux at each simulation
output time in a spherical subvolume above the AR defined as $r \in
\{1R_\odot,3R_\odot\}$, $\theta \in \{0.540,1.915 \}$~rad, and $\phi \in
\{-0.785,+0.785\}$~rad. First, we integrated a large number of magnetic
field lines ($\gtrsim 65~000$) from both the lower and upper boundaries
and tested whether each fieldline is open, i.e., has one end terminate at the
lower boundary and the other at the upper boundary.  Next, we constructed
a pixel mask of the location of open field footpoints (bottom row of
Figure~\ref{FBipoleReconnection}), calculated the radial flux associated
with each open field pixel, and then summed the total to obtain an estimate
of the open flux:
\begin{equation}
\Phi_{\rm op} = \sum B_r(R_\odot,\theta_{j},\phi_{i}) \; \Delta A_{ij} \; , \label{e1}
\end{equation}
where $\Delta A_{ij} = R_\odot^2 \sin{\theta_j} \Delta \theta \Delta
\phi$, $\Delta \theta = 0.0052$~rad, and $\Delta \phi = 0.0061$~rad.
We define the change in open flux as $\Delta \Phi_{\rm op} = \Phi_{\rm
op}(t) - \langle \Phi_0 \rangle$, where $\langle \Phi_0 \rangle = 4.43
\times 10^{21}$~Mx is the mean background open flux in our spherical
sub-volume, taken over from $t=0$ through $t=20~000$~s.

Interchange reconnection conserves the global flux topologies of the
various structures, so the increase seen in the subvolume is offset by
open flux closing down elsewhere in the computational domain, and we
did not count flux associated with field lines that pass through either
of the $r-\theta$ or $r-\phi$ planar boundaries of our subvolume. The
linear decrease in open flux for $t \le 20~000$~s is due to the gradual
reconnection of the overlying field in response to the expansion of
the AR flux system which drives a deformation of the null-point into a
reconnecting current sheet.

Following reconnection onset, the unsheared portion of the AR flux
overlying the southern section of the polarity inversion line is gradually
transferred to the northern half. In addition, the reconnection process
transfers the (globally closed) streamer-belt flux to the southern side of
the AR flux system, shifting the AR flux system ever closer to the global
streamer belt - coronal hole boundary. This closed--closed reconnection
has the effect of both shifting the geometric identity (e.g., particular
set of field lines at any given time) of the AR separatrix boundary
and the outer spine line closer to the open--closed field interface
(e.g., \opencite{Edmondson2009}, \citeyear{Edmondson2010a}). Given our spherical subvolume and the method
used to classify open field lines, this evolution appears as a gradual
decrease of open flux.

The evolution of the open flux has a qualitative resemblance
to the kinetic energy curve. The kinetic energy increases sharply at
$t\sim$18~000~s (indicating the onset of relatively fast reconnection)
whereas the change in open flux lags slightly, starting its rapid rise
at $t\sim$21~000~s. We open approximately $3 \times 10^{20}$~Mx of AR
flux over the 10~000~s it takes for the system to relax enough to start
to close this flux back down. The maximum peak in $E_{\rm K}$ curve is
reached by 22~000~s, while $\Delta \Phi_{\rm op}$ reaches its maximum
peak by 26~000~s.

% ===================================================================
\section{Material, Energy, and Helicity Flux Estimates}
\label{S:Outflow-Flux}
% ===================================================================

We quantified the material, energy, and helicity
fluxes resulting from our IR generated outflow by examining the
fluxes through a radial surface of the spherical subvolume defined in
$\S$\ref{S:Overview-Top} at $r=2R_\odot$.  The total rate of change in
mass, kinetic energy, enthalpy, magnetic energy, and magnetic helicity passing
through this surface is then given by integrating their respective
flux densities,
\begin{eqnarray}
\frac{d M}{dt}   &=& \sum \rho v_r \; \Delta A_{ij} \; , \label{e2} \\
\frac{d E_{\rm K}}{dt} &=& \sum \left( \frac{1}{2} \rho v^2 \right) v_r \; \Delta A_{ij} \; , \label{e3}\\
\frac{d E_{\rm E}}{dt} &=& \sum \left( 5 n k_B T \right) v_r \; \Delta A_{ij} \; , \label{e4}\\
\frac{d E_{\rm M}}{dt} &=& \sum S_r \; \Delta A_{ij} \; , \label{e5}
\end{eqnarray} 
where the sum is over $(\theta_i,\phi_j)$ elements, $\Delta A_{ij}$
is defined in Equation~(\ref{e1}) at $r=2R_\odot$, 
the number density is $n=\rho/m_p$, and $k_B$ is the Boltzmann constant.
The radial component
of the Poynting flux $S_r$ is calculated from $\vec{S}= (4\pi)^{-1}
c \vec{E} \times \vec{B}$ with the electric field $c\vec{E} = -\vec{v}
\times \vec{B}$ given by the advection velocity, yielding,
\begin{equation}
S_r = \frac{1}{4 \pi} \left[ \left( B_\theta^2 + B_\phi^2 \right) v_r - \left( B_\theta v_\theta + B_\phi v_\phi \right) B_r \right] \; .
\end{equation} 
We note that the first term corresponds to the radial transport of magnetic energy associated with the
tangential field components $(B_\theta, B_\phi)$ carried through the surface by $v_r$,
and the second term describes the energy flux associated with
radial fields experiencing lateral transport (see \opencite{Abbett2012} for discussion). In general, attributing components of the magnetic energy flux to well-defined wave-modes (see Section~\ref{S:Waves}) in systems with complex magnetic configurations is not straightforward. However, as we describe below, the largest, positive $S_r$ enhancement corresponds spatially and temporally with the positive outflow of the other material and energy quantities.

We also calculated the flux of relative helicity through our surface
(e.g., \opencite{Berger1984}; \opencite{Berger2000}),
\begin{equation}
\frac{dH}{dt} = 2 \oint_{S} dS \; \left( \vec{A_p} \cdot \vec{v} \right) B_r - \left( \vec{A_p} \cdot \vec{B} \right) v_r  \; .
\label{e7}
\end{equation}
where $\vec{A_p}$ is the vector potential of the reference field (e.g.,
potential field for given boundary conditions). The vector potential
$\vec{A_p}$ was calculated using the flux function, $\vec{A_p}
= \hat{\vec{r}} \times \nabla \psi$ from the $B_r$ distribution at
$r=2R_\odot$. In spherical coordinates, the flux function is given by
\begin{equation}
\psi = -\frac{1}{4\pi} \oint_{S'} dS' \; B_r(\theta',\phi') \ln{\left[\frac{1-\cos{\xi}}{2} \right]}  \; ,
\end{equation}
with $\cos{\xi}$ in the natural logarithm term being the spherical
(angular) distance between $(\theta,\phi)$ and $(\theta',\phi')$ given
by \inlinecite{Berger2000} as
\begin{equation}    
\cos{\xi} = \cos{\theta}\cos{\theta'} - \sin{\theta}\sin{\theta'}\cos{\left(\phi-\phi'\right)} \; .
\end{equation} 
Here, $dS$ and $dS'$ correspond to the same surface area
differentials as in Equations (\ref{e2})--(\ref{e5}). The physical
interpretation of this double integral is the total magnetic-field
rotation at each surface point, due to both the rotation of tangential
flows and twisted flux transported by radial flows.
We note that this calculation is the spherical version of the
Cartesian formalism used to quantify the helicity expulsion in the
\citeauthor{Pariat2009} (\citeyear{Pariat2009}, \citeyear{Pariat2010})
coronal jet simulations.

%
%
%%%%%%%%%%%%%%%%%%%%%%%%%%%%%%%%%%%%%%%%%%%%%%%%%%%%%%%%
%\setcounter{figure}{0}
%\makeatletter 
%\renewcommand{\thefigure}{\@arabic\c@figure a}
%\makeatother
%%%%%%%%%%%%%%%%%%%%%%%%%%%%%%%%%%%%%%%%%%%%%%%%%%%%%%%%
%
\begin{figure}    %%%%%%%%%%%%%%%%%% FIGURE 6 
   \centerline{\includegraphics[width=0.90\textwidth,clip=]{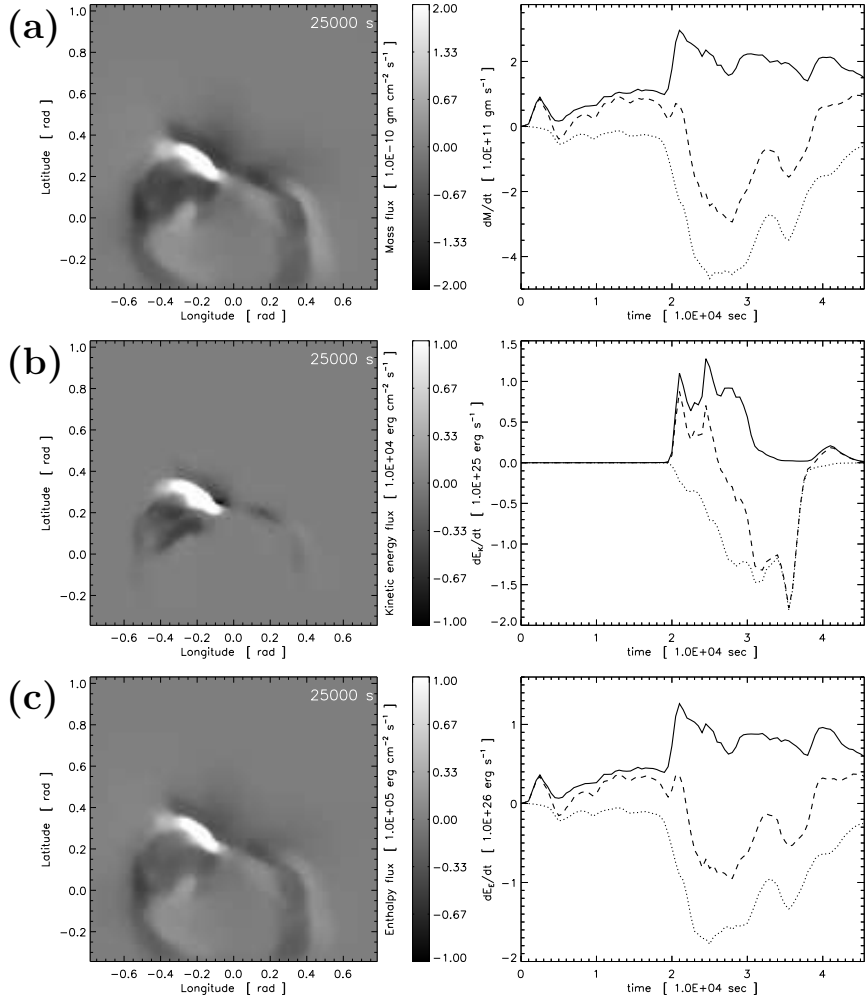}}
   \vspace{-1.10\textwidth}   % Shift close to the panel top 
       \centerline{\Large \bf     
       \hspace{0.0\textwidth}  \color{black}{(a)} \hfill}
   \vspace{0.32\textwidth}    % Shift back to the panel bottom 
       \centerline{\Large \bf      
       \hspace{0.0\textwidth}  \color{black}{(b)} \hfill}
   \vspace{0.32\textwidth}    % Shift back to the panel bottom
       \centerline{\Large \bf      
       \hspace{0.0\textwidth}  \color{black}{(c)} \hfill} 
   \vspace{0.32\textwidth}    % Shift back to the panel bottom 
   \caption{Flux densities through the $r=2R_\odot$ Gaussian surface at $t=25~000$~s:
   (a) mass flux, (b) kinetic energy flux, (c) enthalpy flux. The right column plots the 
   surface integrals corresponding to the time rate of change in mass, kinetic energy, and enthalpy through 
   the radial surface (Equations (\ref{e2})--(\ref{e4})) where the positive fluxes are shown as thick solid lines, negative fluxes as dotted lines,
   and the net flux as dashed lines.
   }
   \label{Ffluxes}
\end{figure}
%
%%%%%%%%%%%%%%%%%%%%%%%%%%%%%%%%%%%%%%%%%%%%%%%%%%%%%%%%
%\addtocounter{figure}{-1}
%\makeatletter 
%\renewcommand{\thefigure}{\@arabic\c@figure b}
%\makeatother
%%%%%%%%%%%%%%%%%%%%%%%%%%%%%%%%%%%%%%%%%%%%%%%%%%%%%%%%
%
\begin{figure}    %%%%%%%%%%%%%%%%%% FIGURE 7 
   \centerline{\includegraphics[width=0.90\textwidth,clip=]{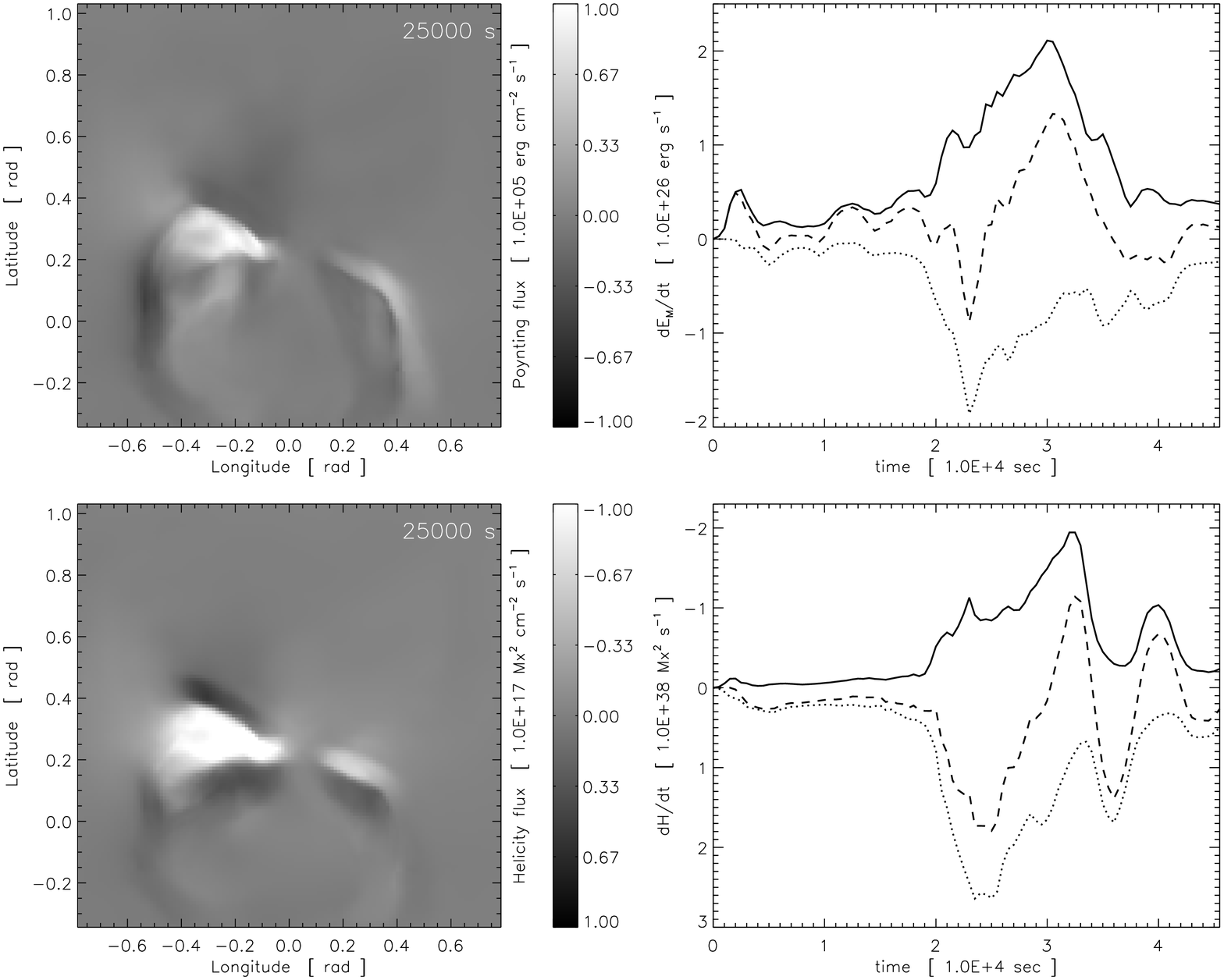}}
   \vspace{-0.725\textwidth}   % Shift close to the panel top 
       \centerline{\Large \bf     
       \hspace{0.0\textwidth}  \color{black}{(a)} \hfill}
   \vspace{0.32\textwidth}    % Shift back to the panel bottom 
       \centerline{\Large \bf      
       \hspace{0.0\textwidth}  \color{black}{(b)} \hfill}
   \vspace{0.32\textwidth}    % Shift back to the panel bottom 
   \caption{Flux densities for the (a) Poynting flux and (b) relative helicity flux 
   in the same format as Figure~\ref{Ffluxes}. The right panels plot the temporal evolution 
   of the surface integral of each quantity (Equation (\ref{e5}) and (\ref{e7})).
   For the total helicity rate the $y$-axis
   values are reversed to align the high flux of negative (left-handed)
   helicity with the reconnection material and energy outflow, thus
   the negative (positive) flux is shown as thick solid (dotted) lines, respectively.
   }
   \label{Ffluxplots}
\end{figure}
%%%%%%%%%%%%%%%%%%%%%%%%%%%%%%%%%%%%%%%%%%%%%%%%%%%%%%%%
%\makeatletter 
%\renewcommand{\thefigure}{\@arabic\c@figure}
%\makeatother
%%%%%%%%%%%%%%%%%%%%%%%%%%%%%%%%%%%%%%%%%%%%%%%%%%%%%%%%

%
Figure~\ref{Ffluxes} plots in rows (a) the mass flux, (b) kinetic energy flux, and (c) the enthalpy flux densities on the
$2R_\odot$ radial surface of our spherical subdomain at the simulation
time $t=25~000$~s in the left panels and the 
temporal evolution of the
surface integral quantities $dM/dt$, $dE_{\rm K}/dt$, and $dE_{\rm E}/dt$ (given by Equations~(\ref{e2})--(\ref{e4})) that represent the time rate of change of each of these quantities passing through the radial surface. 
We have plotted both the positive (negative) material
and energy quantities separately as the thick solid (dotted) lines, respectively. The net totals are shown as the dashed lines.
The linear scaling of the integrated energy plots masks the small but non-zero kinetic energy flux $dE_k/dt$ asocialted with the non-zero mass fluxes $dM/dt$ before the onset of the main phase of IR, but the magnitude is only $\sim 10^{21}$~ergs s$^{-1}$ for $t <$ 20~000~s.

Figure~\ref{Ffluxplots} plots in row (a) the magnetic energy flux density and in row (b) the relative helicity flux density and their surface integral totals (Equations~(\ref{e5}) and (\ref{e7}), respectively) in the same format as Figure~\ref{Ffluxes}.  
For the time
rate of change in magnetic helicity we flipped the $y$-axis values
and show the negative/left-hand helicity flux as the solid line to
facilitate the visual association with the positive material and energy
IR-jet outflow quantities. 

The strong positive flux density enhancements in
mass density, kinetic energy, enthalpy, and magnetic energy flux densities directly
correspond to the interchange reconnection jet outflow.
Likewise, the
strong negative helicity flux density is aligned co-spatially with the
other outflow quantities, and represents the positive radial transport
of left-hand twist on newly opened magnetic field lines (e.g., the
un-twisting of the field stressed by counter-clockwise rotational driving
flows). 
Once the IR starts in earnest at $\sim$20~000~s (visible in Figure~\ref{Feng}),
the outflow signatures appear at $(-0.2, +0.1)$~rad in latitude,
longitude and propagate slightly eastward and towards the northern
pole. 
The outflow flux spatial distributions are complicated after $\gtrsim
30~000$~s because of the signal reflections off of the outer boundary.

We note that there are significant downflows of
both mass and energy present in addition to the expected IR generated
outflows. These are largely due to gravity acting on displaced material. A
back-of-the-envelope calculation of the free fall speeds confirm these
inflows as simply material flowing down recently ``straightened out"
field lines that have either  newly opened or recently reconnected. In the
absence of a background solar-wind outflow, only the material accelerated
by the reconnection jet and carried by the associated waves continue to
propagate radially outwards.

The overall spatial distribution and evolution of the outwardly
propagating mass and energy fluxes are qualitatively similar to the 3D
coronal jet results presented by \inlinecite{Pariat2009}, although as
we discussed previously \cite{Edmondson2009}, this simulation is much
less ``explosive'' because the injected stresses are closer to the AR
separatrix and therefore allows the release of accumulated magnetic energy
through current sheet generation and reconnection much earlier;
only a small amount of the magnetic free energy is released. 
The vast majority of the accumulated relative helicity also remains enclosed in the AR flux system. The relative helicity flux introduced during the uniform phase of the vortical shearing flows at the $r=R_\odot$ lower boundary was $dH/dt = -1.90 \times 10^{39}$~Mx$^2$~s$^{-1}$. Thus, the magnitude of both the positive and negative helicity fluxes through the 2$R_\odot$ surface from the IR process are only on the order of 5--10\% of that associated with the energizing footpoint motions.     

It is also interesting to compare the ratio of maximum values of the integrated
magnetic with kinetic energy fluxes carried by the jet outflow in the
two simulations: the \inlinecite{Pariat2009} results yield a more
kinetic jet $dE_{\rm M}:dE_{\rm K} \sim 4:1$, whereas our results yield
$dE_{\rm M}:dE_{\rm K} \sim10:1$. In both cases the torsional \Alfven
waves provided the bulk of the magnetic energy transport away from the
reconnection region, but the \inlinecite{Pariat2009} jet opens more of the closed flux and is able to transfer 90\% of the total introduced helicity from the closed to open field lines, whereas our IR ``relaxation'' transfers roughly 5\%.
It is also important to recall that for planar \Alfven waves the magnetic
and kinetic energy densities are in equipartition, so our results indicate
the significance of the additional magnetic energy transport associated
with opening up previously closed magnetic flux.

%=====================
%

% ===================================================================
\section{Reconnection-Driven \Alfven Waves}
\label{S:Waves}
% ===================================================================

% ==================
\subsection{Low-Frequency Waves from IR Topological Restructuring}
% ==================

The basic topology and evolution of the generation of large-scale,
nonlinear torsional \Alfven waves associated with interchange
reconnection opening up twisted closed flux are apparent from
Figure~\ref{FBipoleReconnection} as well as in the online animations of
\inlinecite{Edmondson2009}. To characterize the wave properties
of the outwardly propagating disturbances, we  define the magntiude of
the magnetic field fluctuations as
\begin{equation}
\frac{\delta B}{\langle B \rangle} = \frac{|\vec{B}(t) - \langle \vec{B} \rangle_t|}{\langle B \rangle_t}.
\label{efluc}
\end{equation}
Here, our background ``average'' field $\langle \vec{B} \rangle_t$
is computed at a fixed position from the running temporal average of
the magnetic field centered on $t_i$ with an averaging window width
of $2~000$~s. For our simulation output cadence this represents
five data files ($-2\Delta t$ through $+2\Delta t$) and the window
width is similar to the global characteristic \Alfven time
$\langle \tau_A \rangle = 2R_\odot/\langle V_A \rangle$ calculated
in \inlinecite{Edmondson2009}.  In the limit of small fluctuations,
Equation~(\ref{efluc}) reduces to the standard formulation used in
linearized perturbation analysis.

\begin{figure}    %%%%%%%%%%%%%%%%%% FIGURE 8 
   \centerline{\includegraphics[width=1.0\textwidth,clip=]{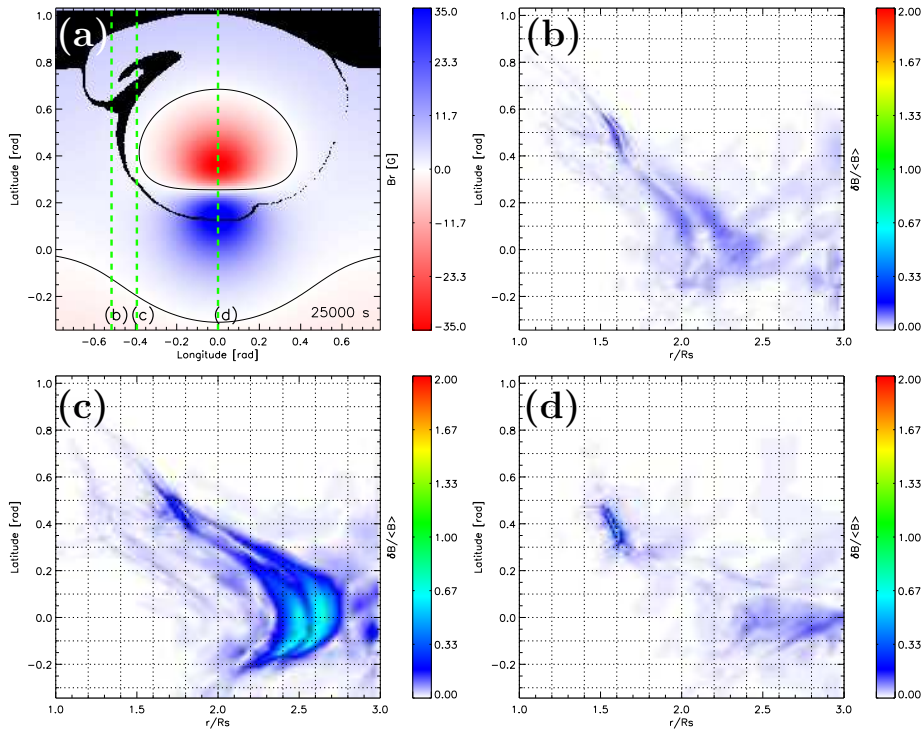}}
   \vspace{-0.81\textwidth}   % Shift close to the panel top 
       \centerline{\Large \bf     
       \hspace{0.040\textwidth}  \color{white}{(a)}
       \hspace{0.415\textwidth}   \color{black}{(b)} \hfill}
   \vspace{0.36\textwidth}    % Shift back to the panel bottom 
       \centerline{\Large \bf      
       \hspace{0.040\textwidth}  \color{black}{(c)}
       \hspace{0.415\textwidth}   \color{black}{(d)} \hfill}
   \vspace{0.35\textwidth}    % Shift back to the panel bottom 
   \caption{Panel (a) shows $B_r$ in the same format as Figure~\ref{FBipoleReconnection}
   with the $\phi$ locations of the three $r$--$\theta$ planes
   indicated with green dashed lines for $t=25~000$~s. Panels (b), (c), and (d) plot
   the normalized fluctuations $\delta B/\langle B \rangle$ that show the spatial
   extent and propagation of the large-scale nonlinear torsional
   \Alfven wave through the domain.
   An animation of this figure is available as an electronic attachment
   to the online version of this article.}
   \label{FdBimg}
\end{figure}

Figure~\ref{FdBimg} plots the $r$-$\theta$ plane of our normalized field
fluctuation quantity $\delta B/\langle B \rangle$ for three longitudinal
values indicated as dashed green lines in panel (a). Panels (b),
(c), and (d) show the planes for $\phi$ values of $\{$-32, -22,
0$\}$~degrees, respectively. These planar cuts sample the open field
corridor that initially opens on the eastern side of the AR separatrix
dome. The animation of this figure (included as an electronic supplement to the online version of
this article) shows the formation and propagation of the torsional wave,
illustrated most clearly in panel (c). The twisted/helical structure
of the propagating wave packet, i.e. strong regions of oppositely
directed $\delta B_\phi$, creates the dual-peak signature in the
fluctuation magnitude plots. One can compare the $\delta B/\langle
B \rangle$ visualization here with the material and energy fluxes in
Section~\ref{S:Outflow-Flux} (Figure~\ref{Ffluxes}), which corresponds to
the midpoint of the $x$-axis range. The flux evolution of $E_{\rm K}$,
$E_{\rm M}$, and $H_r$ can be seen as the intersection of the torsional
\Alfven wave passing through the $r=2R_{\odot}$ radial surface.

\begin{figure}    %%%%%%%%%%%%%%%%%% FIGURE 9 
   \centerline{\includegraphics[width=1.0\textwidth,clip=]{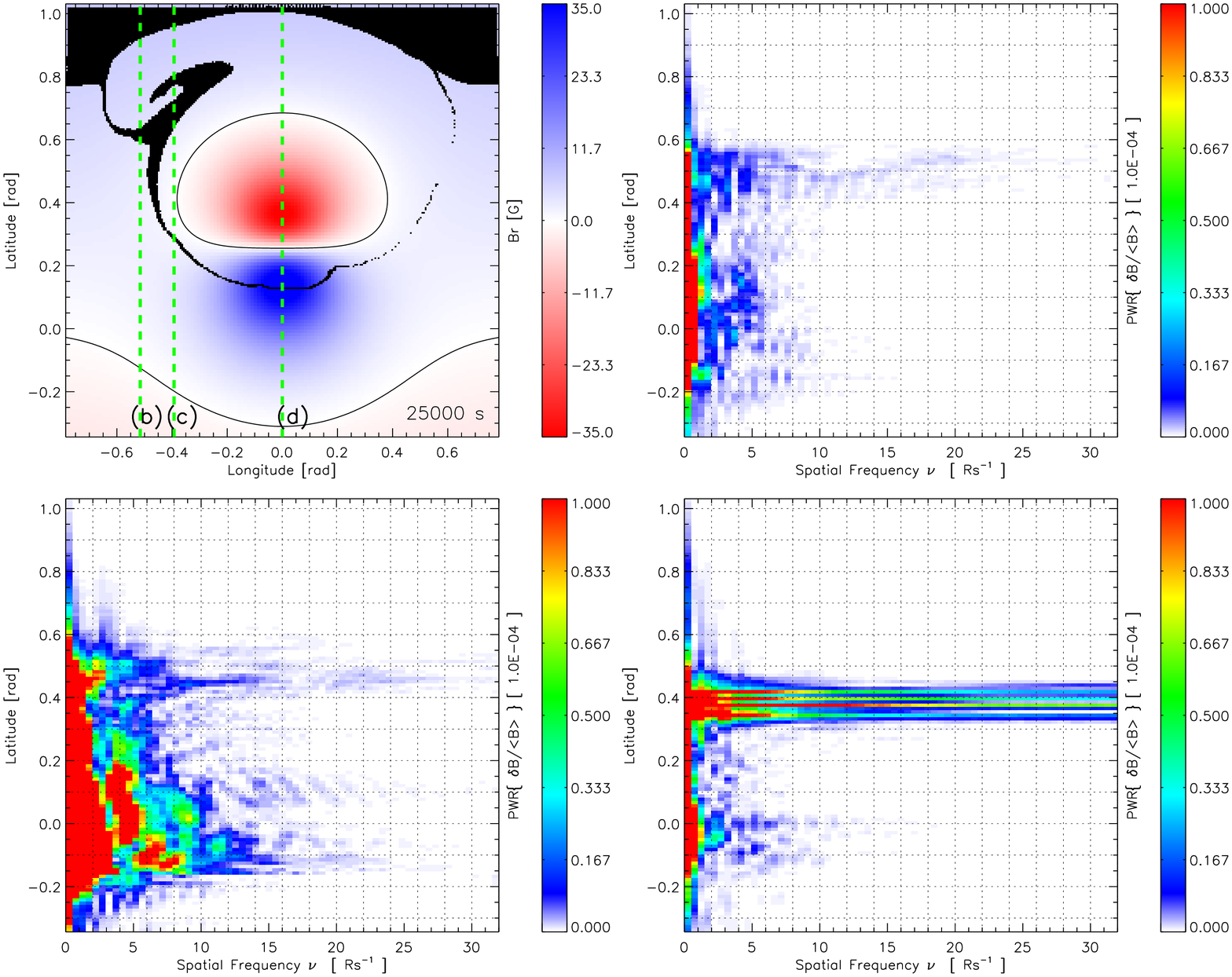}}
   \vspace{-0.81\textwidth}   % Shift close to the panel top 
       \centerline{\Large \bf     
       \hspace{0.040\textwidth}  \color{white}{(a)}
       \hspace{0.415\textwidth}   \color{black}{(b)} \hfill}
   \vspace{0.36\textwidth}    % Shift back to the panel bottom 
       \centerline{\Large \bf      
       \hspace{0.040\textwidth}  \color{black}{(c)}
       \hspace{0.415\textwidth}   \color{black}{(d)} \hfill}
   \vspace{0.35\textwidth}    % Shift back to the panel bottom 
   \caption{Panel (a) again indicates $B_r$ and the open flux
   regions. Panels (b)--(d) show the power spectra of $\delta
   B/\langle B \rangle$ depicted in Figure~\ref{FdBimg}. The sharp,
   filamentary changes in the field associated with the separatrix
   evolving correspond to power across a broad range of $\nu$ 
   including the highest spatial frequencies, whereas
   the propagating, large-scale torsional wave are clearly visible
   as low spatial frequency enhancements. An animation of this
   figure is available as an electronic attachment to the online
   version of this article.}
   \label{FPWRdBimg}
\end{figure}

After visualizing the coherent wave packet generation and propagation
in our 3D MHD simulation data, we now examine the geometric extent
and spectral properties of the wave train associated with our idealized
``discrete'' IR event.
We expect some aspects of the spatial size of the $\delta B/\langle
B \rangle$ signal to be related to the geometry of our system --
specifically the size of the energized AR separatrix, as well as
the global magnetic-field structure. As seen in Figure~\ref{FdBimg},
the central region of the propagating twist wave (panel c) is between
latitude values of $\left[-0.2, +0.2\right]$~rad. Here, the double
peaks of the $\delta B /\langle B \rangle$ magnitude structure each
have a characteristic width of approximately $\lambda_r \sim 0.2 -
0.25 R_\odot$ corresponding to a full wavelength in $\delta B_\phi$
of $\lambda_r \sim 0.4 - 0.5 R_\odot$.

Figure~\ref{FPWRdBimg} plots the normalized fluctuation power as a
function of the radial spatial frequency (simply inverse wavelength) $\nu
= \lambda_r^{-1}$ at $t=25~000$~s for each of the three $r$--$\theta$
planes in Figure~\ref{FdBimg}. Here, panels (b)--(d) plot the
$\theta$-distribution of the $\delta B/\langle B \rangle$ power
spectra. The most common characteristic of power spectra of all three
planar cuts is that for the most part, the relatively strong power is in
the low spatial frequencies (longer wavelengths), demonstrating that the
wavelengths generated by this isolated IR process are relatively long. An
animation of Figure~\ref{FPWRdBimg} showing the temporal evolution of
the normalized power spectra is included as an electronic attachment to
the online version of the article.

Upon closer examination, the power spectra through all three planar
cuts exhibits a bimodal distribution in latitude, with strong power
structures extending through the lower spatial frequencies, peaking in
the angular ranges $\left[ -0.2 , +0.2 \right]$~rad, and $\left[ +0.4 ,
+0.5 \right]$~rad, respectively. Physically, the bimodal distribution
reflects the planar cuts sampling the different field topologies: open
coronal hole, globally closed streamer belt, and locally closed AR. The
relatively narrow, higher latitude peaked distribution reflects the
reconnection-driven wave patterns directly related to the reconnection
site, and opening of the coronal hole channel and the generation of
the open field ``thumb'' structure on the northwest side of the AR
(clearly seen in panel (a)). We note that there is a systematic shift of
approximately 0.1 radians to lower latitudes with respect to the open
field footpoint pattern of panel (a), due to the global magnetic geometry
of the streamer belt - coronal hole pattern acting as an equatorward
wave-guide for the system dynamics above the surface. On the other hand,
the relatively broad, lower latitude peaked distribution reflects
the reconnection-driven wave patterns of the global restructuring of
the streamer belt, as the AR shifts into the coronal hole.

The strongest power exhibited in all three planar cuts is at the lowest
spatial frequency $\nu \lesssim 1$, suggesting that the dominant wavelength
is directly related to to the size of the separatrix dome; estimated
by the angular spread between footpoints plotted in panel (a), the
widest part at latitude $\sim$0.4, the  angular width stretches from
approximately $-0.5$ through $+0.5$ in longitude, corresponding to a characteristic
size of $\sim 1 R_{\odot}$. We note that this is only an estimate and does
not account for height or writhe, which serve to increase the length of
newly reconnected field, decreasing the corresponding spatial frequency.

The strong power in the lower latitude range, $\left[
-0.2 , +0.2 \right]$~rad has a broad angular extent and remains
in the lower spatial frequency range very simply due to the global size
of the streamer belt. The upper boundary of this distribution, latitude
of $+0.2$~rad is approximately the lower extent of the AR structure. Thus,
the broad power reflects the restructuring of the streamer belt as the AR
shifts into the coronal hole, generating strong wavelength patterns in the
range of order the size of the AR down to the length of the reconnecting
current sheet; all three panels offer evidence that this lower power
distribution broadly extends to the same spatial frequency level at the
upper power distribution. That the strongest waves are shown in panel (c)
reflects the fact that the driving flow is counter-clockwise, causing the
IR process to move in a western direction (see \opencite{Edmondson2009}
for details), generating the bulk of the streamer-belt restructuring
wave activity on the western limb of the AR structure.

The large-scale torsional \Alfven wave signal, most visible in panel (c),
associated with the coronal hole boundary jump across the AR, generates
the wave power feature at approximately $\nu \lesssim 7/R_{\odot}$ at
latitude $-0.1$~rad, as well as the clear extended peak in latitude
around approximately $\nu \sim 4/R_{\odot}$ between latitudes $(-0.05,
+0.2)$~rad.
We note that the range of $\delta B/\langle B \rangle$ spatial scales
associated with our interchange reconnection wave activity ($7/R_{\odot}
\lesssim \nu \lesssim 2/R_{\odot}$; 100~Mm~$\lesssim \lambda_r \lesssim$
350~Mm) fall within the range of the nonturbulent periodic density
structures observed in the STEREO/SECCHI data remotely and in the
solar wind in-situ \cite{Viall2010}, as well as comfortably within the
low-frequency range of the interplanetary \Alfven wave spectra (e.g.,
\opencite{Belcher1971}).

% ==================
\subsection{High-Frequency Waves from the IR Current Sheet}
% ==================

In all three Figure~\ref{FPWRdBimg} planar cuts, there is also strong power in the upper
latitude range, $\left[ +0.4 , +0.5 \right]$~rad, extending to higher
spatial frequencies, relative to the lower latitude distribution. 
This effect reflects a direct connection to the reconnection site that generates waves at higher spatial frequencies.
From the corresponding panels in Figure~\ref{FdBimg}, the spatial scale of
the reconnection site (located in the range $r \sim 1.5 - 1.7 R_{\odot}$ and
co-latitude $\sim 0.4 - 0.5$~rad depending on the panel) is estimated to
be $L \sim 0.2 - 0.3 R_{\odot}$ corresponding to a spatial frequency of
$\nu \sim 7.5 / R_{\odot}$; the effective upper spatial frequency limit of
the relatively broad upper latitude strong power distribution of panels
(b) and (c). In each panel, there is evidence of much narrower power
distribution extending to even higher spatial frequencies. This effect
is a remanent of even smaller scale structures within the reconnecting
current sheet itself, down to the grid scale of the simulation. In fact,
such high spatial frequency effects are abundant in panel (d), since
this plane directly cuts the reconnection current sheet, and therefore
samples the smallest scale structures of the simulation. 

We note that because highest frequency waves originate from reconnection sites, the computational grid scale and the magnetic resistivity employed place an artificial upper limit on the wave frequencies that we are able to resolve in the simulation. The tearing and breakup of currents sheets generate a whole distribution of magnetic islands, and the associated high-frequency waves generated will have characteristic wavelengths on the order of the size of the islands \cite{Drake2006,Isobe2008,Ji2011,Barta2011,Shen2013}. 
Much higher resolution simulations will be needed to resolve the higher frequency wave modes. Recent studies of current sheet formation and reconnection with adaptive grid refinement \cite{Edmondson2010b,Karpen2012} are beginning to resolve the current sheet substructure and island formation in global MHD models while employing a ``numerical resistivity'' model. However, to characterize the highest frequency wave components it will be important to investigate the effects of the magnetic resistivity model on the evolution of the current sheet evolution and reconnection. We have focused on the low-frequency torsional Alfv\'{e}nic wave activity because our large-scale system evolution should be relatively independent of the specific details of the magnetic reconnection. All we require here is that some moderate amount of previously closed flux that has accumulated a significant twist/shear component is allowed to become open.

%
% ==================

%
% ===================================================================
\section{Summary and Discussion}
\label{S:Discussion}
% ===================================================================

We have presented a detailed analysis of the \inlinecite{Edmondson2009}
MHD simulation of closed-to-open interchange reconnection in an idealized
AR flux system at the edge of the helmet streamer-belt.
Despite the limitations of the simulation, i.e., the extremely large
AR source and laminar vortical energization flows, our time-integrated
IR-driven Poynting flux carries only $\sim$1\% of the accumulated free
magnetic energy introduced into the system yet generates significant
coronal dynamics; notably the generation of a large-scale, torsional
\Alfven wave that transmits the shear and twist components of the
energized, previously closed AR flux into the open field region.
We visualized this large-scale wave activity as a succession of field-line plots (Figure~\ref{FBipoleReconnection}), as well as the formation and propagation of the $\delta B/\langle B \rangle$ structure (Figure~\ref{FdBimg} and its online animation).
One of our primary goals was to emphasize the intrinsic
and fundamental connection between reconnection and \Alfven-wave
generation.

Large-scale torsional \Alfven waves have been identified in previous
simulation work on coronal jets \cite{Shibata1986,Pariat2009} and flux
emergence \cite{Torok2009}, and appear to explain recent observations
that show evidence of helical structure or apparent twisting/un-twisting
motions \cite{Patsourakos2008,Liu2009,Kamio2010,Shen2011}, and may
be a significant component of the ubiquitous wave activity observed
(e.g., \opencite{Cirtain2007}). 
Observations have shown that impulsive, jet-like processes are operating over a wide range of spatial scales in both open- and closed-field regions and in different layers of the atmosphere, including photospheric and chromospheric jets and spicules \cite{DePontieu2007,Sterling2010,Liu2011a}, H$\alpha$ surges \cite{Kurokawa1993,Canfield1996}, low coronal jets in EUV \cite{Chae1999,Jiang2007,Patsourakos2008} and X-rays \cite{Shibata1997,Alexander1999,Cirtain2007}. 
\citeauthor{Liu2011a} (\citeyear{Liu2011a,Liu2011b}) have measured both the ÒstationaryÓ component where the entire jet appears to move back and forth, which is indicative of transverse footpoint motion, as well as the impulsive, outwardly propagating wave transients driven by reconnection -- and similar features have been reproduced in numerical simulations of \cite{MorenoInsertis2008,Murray2009}.
At all spatial scales, magnetic reconnection processes drive \Alfven waves into the corona and in unipolar open field regions, ultimately into the solar wind.

Recently, \inlinecite{Cranmer2010} constructed a kinematic model that
included both flux emergence and quasi-static evolution to quantify
the energy fluxes generated by IR process and concluded that IR has
trouble producing values within the range of ${\rm a~few} \times
10^5$--$10^6$~erg~cm$^{-2}$~s$^{-1}$ needed to accelerate the solar
wind. 
As shown in Figure~\ref{Ffluxes}, our calculation of the radial
Poynting flux through a Gaussian surface above the IR site readily
produces $10^5$~erg~cm$^{-2}$~s$^{-1}$ for the duration of the dynamic
flux opening ($\sim$2.5~hours). The radial Poynting flux carried by our singular interchange reconnection event ``jet'' subtends an angular area of $\sim$0.20\% of the $4\pi$ spherical surface. However, our simulation probably represents the largest conceivable scale IR processes operate on before being considered a slow solar wind ``streamer blob'' and the fluctuations of these wavelengths are likely to escape directly into the solar wind without contributing much to the heating process. 

To argue that IR could supply {\textit{all}} of the necessary power to heat the corona and accelerate the solar wind, one would have to (1) determine how the IR-driven \Alfven wave properties scale with bipolar flux system properties (field strength, flux content, geometric size) and energization mechanism (rotational and translational motions, footpoint shuffling associated with granular and super-granular diffusion), and (2) estimate the distribution of bipolar flux system source sizes from high-resolution coronal magnetic-field extrapolations. 
If IR were a continuous stochastic, intermittent process and
the IR-driven wave power and characteristic wavelengths reflected their
source region structure, then one could estimate this contribution to
the low-frequency portion of the observed interplanetary \Alfven wave
spectra.
Since the topological structure of nested flux systems with a complex
network of separatrices and quasi-separatricies (favorable sites
of IR) is exactly what high-resolution PFSS extrapolations and the
resulting Q-maps show \cite{Linker2011,Titov2011} and the basis of the
\inlinecite{Antiochos2011} S-Web model -- we are looking forward to
future progress in this arena.

\begin{acks}
BJL and YL acknowledge support from the AFOSR YIP FA9550-11-1-0048,
NASA HTP NNX11AJ65G, and HGI NNX08AJ04G. JKE acknowledges support
NASA LWS NNX10AU57G and NNX07AB99G.
\end{acks}

%%%% BIBLIOGRAPHY %%%%%%%%%%%%%%%%%%%%%%%%%%%%%%%%%%%%%%%%%%%%%%%%%%%%%%%%%%%
%\mbox{}~\\ 
%\noindent {\normalsize \bf Bibliography Included with \BibTeX }\\* 
%      % more powerful
%  With \BibTeX\ the formatting will be done automatically for all 
%the references cited with one
%of the \verb+\cite+ commands (Section~\ref{S-references}).
%Besides the usual items, it includes the title of the article 
%and the concluding page number. 
%   
%     % format of references provided by the journal (.bst)
%\bibliographystyle{spr-mp-sola}
\bibliographystyle{spr-mp-sola-cnd} %% Alternative style: no title,
%                                      % no concluding page. 
%
%     % name your Bibtex file containing your references (.bib)
\bibliography{str2_v4_arxiv}  
\end{article} 
\end{document}